\documentclass[journal=nalefd,manuscript=article]{achemso} 

\usepackage{amsmath}  \usepackage{amssymb}  \usepackage{amsfonts}  \usepackage{bm}  
\usepackage{graphicx}  
\usepackage{xcolor}
\usepackage{float}

\usepackage{xcolor}

\title{High efficiency coupling of  free electrons to sub-$\lambda^3$ modal volume, high-Q photonic cavities}

\author{Malo Bézard}
 \affiliation{%
 Universit\'e Paris-Saclay, CNRS, Laboratoire de Physique des Solides, 91405, Orsay, France
}%
\author{Imene Si Hadj Mohand}%
\affiliation{%
 Centre de Nanosciences et Nanotechnologies, CNRS, Université Paris Saclay, 10 Bd Thomas Gobert, 91120 Palaiseau}

\author{Luigi Ruggierio}
\affiliation{%
 Universit\'e Paris-Saclay, CNRS, Laboratoire de Physique des Solides, 91405, Orsay, France
}%

\author{Arthur Le Roux}
\affiliation{%
 Centre de Nanosciences et Nanotechnologies, CNRS, Université Paris Saclay, 10 Bd Thomas Gobert, 91120 Palaiseau}

\author{Yves Auad}
\affiliation{%
 Universit\'e Paris-Saclay, CNRS, Laboratoire de Physique des Solides, 91405, Orsay, France
}%

\author{Paul Baroux}
\affiliation{%
 Centre de Nanosciences et Nanotechnologies, CNRS, Université Paris Saclay, 10 Bd Thomas Gobert, 91120 Palaiseau}
\author{Luiz~H.~G.~Tizei}
\affiliation{%
 Universit\'e Paris-Saclay, CNRS, Laboratoire de Physique des Solides, 91405, Orsay, France
}%

\author{Xavier Checoury}
\email{xavier.checoury@c2n.upsaclay.fr}
\affiliation{%
 Centre de Nanosciences et Nanotechnologies, CNRS, Université Paris Saclay, 10 Bd Thomas Gobert, 91120 Palaiseau}

\author{Mathieu Kociak}
 \email{mathieu.kociak@universite-paris-saclay.fr}
\affiliation{%
 Universit\'e Paris-Saclay, CNRS, Laboratoire de Physique des Solides, 91405, Orsay, France
}%

\begin{document}
\begin{abstract}
We report on the design, realization and experimental investigation by spatially resolved monochromated electron energy loss spectroscopy (EELS) of high quality factor cavities with modal volumes smaller than $\lambda^3$, with $\lambda$ the free-space wavelength of light. The cavities are based on a slot defect in a 2D  photonic crystal slab made up of silicon. They  are optimized for high coupling of electrons accelerated to 100 kV, to quasi-Transverse Electrical modes polarized along the slot direction. We studied the cavities in two geometries. The first geometry, for which the cavities have been designed, corresponds to an electron beam travelling along the slot direction. The second consists in the electron beam travelling perpendicular to the slab. In both cases, a large series of modes is identified. The dielectric slot modes energies are measured to be in the 0.8- 0.85 eV range, as per design, and surrounded by two bands of dielectric and air modes of the photonic structure. The dielectric even slot modes, to which the cavity mode belongs, are highly coupled to the electrons with up to  3.2$\%$ probability of creating a slot photon per incident electron.  Although the experimental spectral resolution (around 30 meV) alone does not allow to disentangle cavity photons from other slot photons, the remarkable agreement between the experiments and finite-difference time-domain (FDTD) simulations allows us to deduce that amongst the photons created in the slot, around 30$\%$ are stored in the cavity mode.  A systematic study of the energy and coupling strength as a function of the photonic band gap parameters permits to foresee increase of coupling strength by fine-tuning phase matching. Our work demonstrates free electron coupling to high quality factor cavities with low mode density, sub-$\lambda^3$ modal volume, making it an excellent candidate for applications such as quantum nano-optics with free electrons.  
\end{abstract}

\maketitle


Free electron based spectroscopies, such as electron energy loss spectroscopy (EELS), photon induced near field electron microscopy (PINEM) or cathodoluminescence (CL) have been extensively used to investigate optical properties of nano-objects \cite{Polman2019}. One of the main drives of early studies was the impressive spatial resolution of these techniques, with sub-nanometer resolution in the (scanning) transmission electron microscopes ((S)TEM) or few nanometer resolution in scanning electron microscopes (SEM). Given the very high localization of electromagnetic fields in plasmonic nanoparticles, it is no surprise that the first studies of CL \cite{Yamamoto2001}, EELS \cite{Nelayah2007,Rossouw2013}, and PINEM \cite{Barwick2009} focused on them. With the increase of the spectral resolution in EELS and improvement in signal-to-noise ratio for CL, excitations with increasingly higher quality factors (Q) excitations could be studied, should they be surface plasmons \cite{Rossouw2013,Saito2015}, surface phonons \cite{Lagos2022} and hybridized modes \cite{Tizei2020}, or photonic excitations \cite{Hyun2008,Sapienza2012,Auad2022}. 

However, orders of magnitude improvement (from typically $10-100$ to $10^5$ or higher) is needed to access the high quality factors (high-Q) cavities necessary to explore quantum excitations with fast electrons \cite{DiGiulio2019,Kfir2019}. Also, a high coupling between the electron and the cavity is required \cite{GarciiaDeAbajo2013}. Finally, a low modal volume is extremely pertinent for quantum nano-optical experiments in general, and well adapted to the possibilities offered by the excellent spatial resolution of the electrons.

As the linewidths of high-Q cavities (typically in the $\mu$eV range in the near-infrared) are intrinsically much smaller than the energy resolution of state-of-the-art EELS systems (at best a few meV \cite{Krivanek2009}), other alternatives have been considered in the past years for measuring their quality-factor.  Indeed, using tunable lasers in conjunction with PINEM, electron energy gain spectroscopy (EEGS) became recently possible, unravelling the predicted \cite{Abajo2008a} promise of a laser spectral resolution with the spatial resolution of an electron \cite{Wang2020,Henke2021,Auad2023}.

At the time being only few types of cavities, including  defects in plasmonic materials \cite{Saito2015,Saito2019}, spheres subtending whispering-gallery modes \cite{Hyun2008,kfir2020,Auad2022, Auad2023} or ring resonators \cite{Henke2021} have been investigated with electron beam techniques. These cavities suffer from either relatively low Q (less than a thousand) and/or large modal volumes. Among the alternatives, cavities made up of defects in photonic band gap (PBG) materials are very attractive. Indeed, in the case of a bidimensional photonic-crystal slab  made up of a lattice of cylindrical holes (hole slab \cite{Joannopoulos2008}) in a high index material, removing one or several holes creates a highly spatially localized, low modal volume, cavity mode with energy within the band gap of the photonic crystal. A pioneer work on EELS of such a $Q \lesssim 2.10^4$ PBG-based cavity was reported \cite{LeThomas2013}. Unfortunately, the geometry of the cavity did not allow for either high coupling nor for high signal to noise ratio. In these cavities, high quality factors can be achieved by shifting the holes positions close to the created defect. This smooths the index contrast between the cavity and the rest of the crystal \cite{Akahane2003}. It therefore prevents the defect mode from scattering too much towards the vertical direction, avoiding a major source of energy dissipation. This allows to attain record Q/V values, where V is modal volume,  and Q larger than $10^6$ \cite{Akahane2003,Han2010}. Efficient coupling of such cavities to electron beams has never been considered. Therefore, the demonstration of a defect in a PBG design suitable for free electron coupling is still to be done.


In this letter, we investigate the highly monochromated spatially-resolved EELS response of  cavities based on a slot inside a 2D photonic-crystal hole slab made up of  silicon. The slot is replacing a (missing) row of holes leading to the formation of new slot modes. Few holes close to the slot are slightly shifted in order to create a high-Q, sub-$\lambda^3$ modal volume cavity. The design is such that an electron beam passing in the center of the structure does not experience elastic scattering despite travelling over microns. The cavity mode is based on a dielectric  quasi-TE slot mode, and designed such that the field is polarized along the beam path and maximum in the middle of the slot (even mode).  Such an engineering aims at  efficiently increasing the coupling of the electron to the field mode. In order to investigate the structure and identify the different modes observed in the EELS, we have coupled experimental investigations along two main directions, parallel and perpendicular to the slot, to finite-difference time-domain (FDTD) simulations. The dielectric slots modes are experimentally identified. Their measured field distribution, in both parallel and perpendicular directions, corresponds to the expected dielectric quasi-TE modes. Their energy lie between 0.80 eV and 0.85 eV for hole diameters from 180 nm to 280 nm and a 426 nm spacing. It is, as per design, within the band gap of the photonic-crystal slab. In the parallel direction, a high coupling results in a photon creation probability of more than 1 per 30 electrons in the dielectric even slot modes. The other main  groups of modes are identified and mapped, especially those related to the bottom (dielectric) band and upper (air) bands of the photonic-crystal slab. The energy resolution of the experiment does not allow to measure the Q factor of the cavity and to disentangle the cavity mode from the other dielectric slot modes. However, it is sufficient to isolate dielectric slot modes from the air slot modes  as well as from the slab modes. Simulations permit  to estimate the spectral weight of the cavity mode compared to that of other dielectric slots modes. Comparison of experimental results and simulations shows that approximately up to 1 photon is created in the cavity every 100 incoming electrons thanks to our design optimization. Finally, we investigated the energy and intensity behaviour of the cavity modes as a function of the hole size as a way to tune the phase matching condition between the electron and the modes. Therefore,  our experiments supported by  simulations show high coupling values, that can be further enhanced, to high-Q, sub-$\lambda^3$ modal volume cavities. This paves the way to efficient study of quantum optics with free electrons with photonic-crystal-based cavities.


As introduced, the  design of a  high-Q, low modal volume cavity optimized for coupling to free electrons has to rely on a few requirements. First, a long interaction path for the electrons, in the spirit of dielectric laser accelerators \cite{England2014}, is needed. This favours 1D cavities designs over 0 D ones previously studied by EELS \cite{LeThomas2013}. 1D defects can be created by removal or addition of holes series in an air slab photonic-crystal for example. The electrical field of the cavity mode  needs to be aligned with the 1D defect direction, so that an electron travelling in-plane along the defect can efficiently couple to it, and its intensity must be located in the center of the 1D defect. The defect must be transparent to electrons (i. e., a slot \cite{DMello2023}), so that the latter do not undergo any elastic or bulk inelastic scattering, in contrast to an opaque design for the cavity \cite{LeThomas2013}. Last but not least, the design must allow efficient phase matching.

\begin{figure}[H]
\includegraphics[width=14 cm]{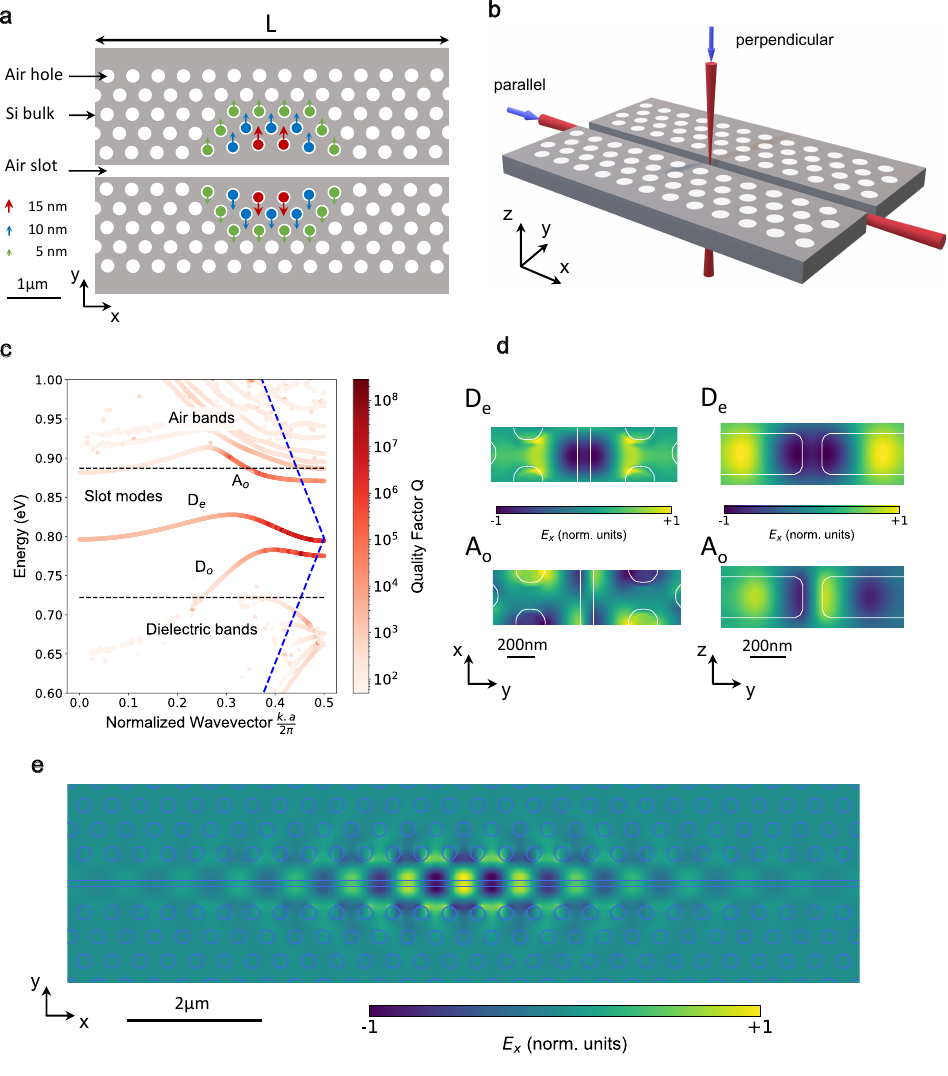}
\caption{\label{fig:design} \textbf{Design of the EELS-compatible low modal volume and high-Q cavity}. a. The cavity is made up of a slot separating two 2D hole-slabs photonic crystals, with slight shifts of the holes indicated by coloured arrows. b. Although designed to be used with electrons travelling parallel to the slot, we have also prepared samples that can be investigated with electrons travelling perpendicular to it, as exemplified here with the two red cones schematizing the propagation of the electron beam in the two geometries. c. Band diagram of the structure for the TE even modes. The band diagram has been calculated for $\rho = 0.255$. Different modes are enlightened. Bottom of the air bands and top of the dielectric bands define the band gap and are indicated by black dashed lines.   The electron dispersion relation is presented as blue dashed lines and calculated for an accelerating voltage of 100 kV. d. Simulations for the $E_x$ electrical field for the modes $D_e$ and $A_o$ taken in the mid-plane of the slab. e. Simulation for the $E_x$ electrical field of the high-Q cavity mode taken in the mid-plane of the slab.}
\end{figure}

As shown in Figure \ref{fig:design}a, we have achieved such a design, consisting in a slot
separating two halves of an hexagonal lattice of air cylinders within a slab of silicon.
Here, the slab is 220 nm thick, the lattice parameter is a = 426 nm and the cylinder radii r = 0.255 a. The slot width is 100 nm. The waveguide formed by replacing one row of air holes by the slot has been enlarged by shifting all the holes so that the two halves are separated by $1.2a\sqrt{3}$, realizing what is called a $W1.2$ waveguide in the photonic crystal literature. These parameters have been optimized to allow high coupling with electrons at around 100 kV. We define the $x$, $y$ and $z$ directions to be along the slot, perpendicular to the slot in the slab plane, and perpendicular to the slab plane (see Figure \ref{fig:design}a and b). Although we have designed the cavity to be used with electrons travelling along $x$ (parallel direction in Figure \ref{fig:design}b), we also studied samples for electron travelling along or close to the $z$ direction (perpendicular direction in Figure \ref{fig:design}b) to confirm the consistency between the design and experimental realization.


To understand the formation of cavity modes and how they couple to free electrons, we start by considering an hypothetical infinite slot defect in a hexagonal air slab. The photonic-crystal slab possesses quasi-TE and quasi-TM modes ("quasi" refers to the fact that although the field is essentially in the mid-plane of the slab, some $z$ component exists due to the finite thickness of the slab \cite{Joannopoulos2008}). For simplicity, those modes will  be referred to as TE and TM in the following. The TE  dispersion relation is shown on Figure \ref{fig:design}c, and the TM one in the Supplementary Figure S1 (see Methods for simulation details).  The former are field-symmetric around the $z=0$ plane (see Figure \ref{fig:design}d and Supplementary Figure S2), and are polarized so that the $E_z$ component is negligible compared to the in-plane components. The opposite holds for TM modes. Only the TE polarization band structure possesses a band gap. Finally, since TM modes are not well localized and are phase matched  with 100 kV electrons only at energies above the TE band gap (see supplementary Figure S1), they will not be considered further.


The slot defect acts as a waveguide for TE modes, adding several modes within the band gap as shown in the Figure \ref{fig:design}c.
For symmetry reasons, the guided TE modes may be even or odd with respect to the y = 0 plane (see Figure \ref{fig:design}d). The modes with the lowest energies are  even and odd TE modes. They lie well within the band gap (see Figure \ref{fig:design}c), and act as dielectric modes \cite{Joannopoulos2008} with their $E_x$ field intensities peak in-between with the photonic-crystal holes (see Figure \ref{fig:design}d and supplementary Figure S2). They are coined as $D_e$ and $D_o$ in Figure \ref{fig:design}.  The modes with higher energies ($A_o$ and $A_e$, the latter not shown on Figure \ref{fig:design}c) arise already very close to the upper band and are air bands, i. e. their $E_x$ field intensities is shifted by half a period with respect to the dielectric modes (see Figure \ref{fig:design}d and supplementary Figure S2). As expected,  the four TE slots modes exhibit  $E_z$ values that are typically one order of magnitude smaller than  $E_x$, see supplementary Figure S2. 

TE modes are easily distinguishable from the photonic-crystal modes, they are well localized inside the slot where the electron can  pass, and they are essentially polarized along the $x$ direction. Therefore, they are good candidates for cavity modes. Also, close to the end of the Brillouin zone, they exhibit large Q factors (Figure \ref{fig:design}c).  Finally, the $D_o$ modes have a null x component of the electrical field in the center of the slot, while the $D_e$ ones have their $x$ component maximum right in the center of the slot. The latter are then easier to target with an electron beam, and will therefore form the basis of the cavity modes.

In a realistic, finite-size design, the  TE modes dispersion curve get discretized. In this case, TE modes form Fabry-Perot (FP) modes of increasing orders. The FP mode separation scales as $a/L$, where $L$ is the cavity length.

 To increase the quality factor, some of the central holes are shifted by few nm \cite{Kuramochi2006}, see Figure \ref{fig:design}a. By doing so, a mode of slightly lower energy is forming the desired cavity mode. In the present study, the effective length of the cavity is only a few periods in length, resulting in a modal volume smaller than  $0.1\lambda^3$ as deduced from simulations, where $\lambda$ is the wavelength in vacuum.  

 The length of the cavity can be monitored by changing the number of displaced holes. In this paper, we have essentially looked at short structures described in Figure \ref{fig:design}a. On both sides of the shifted holes along $x$, a constant number of 17 holes delineates the whole structure, so that the total length of the structure is   15.25 $\mu$m. It is noted that we have tried several other structures with higher number of shifted holes (total length from 22.5 $\mu m$ to 48.5 $\mu m$). However, in these cases, it was practically impossible to perform hyperspectral imaging on the cavity due to the strong charging of the dielectric slab induced by the electron beam illumination.  The Figure \ref{fig:design}b presents the two main geometries used in this paper, either parallel to the slot  or perpendicular to it. Due to the high thickness of silicon to be traversed in the perpendicular geometry, not to mention the parallel geometry, the missing row area was removed allowing a free propagation of the electrons through a slot. Of course, two different sets of samples, as presented in Figure \ref{fig:design}b have been prepared, because there is no possibility of tilting the same sample over 90$^{\circ}$. In practice, the samples have been prepared by electron lithography as described in the Methods. Twin structures with the same design have shown Q-factors of the order of $2.10^5$.

EELS experiments have been performed on a NION Hermes microscope (CHROMATEM) at 100 keV, fitted with a Merlin Quantum Detector camera. Data analysis has been performed using  Hyperspy\cite{francisco_de_la_pena_2022_7263263}, see Methods for details. As described in the Methods, FDTD simulations have been performed for simulating spectra and maps. Simulated spectra have been convolved by a  Gaussian with 40 meV full width at half maximum to take into account experimental broadening. Contrary to experiments where the maps are created by integrating the spectral signal over a given energy window, simulation maps have been drawn for individual modes only in simulations. 





In the following experimental and simulation results, two parameters are of interest, namely the lattice parameter $a$, and the radius $r$ to lattice parameter ratio $\rho= r/a$ . The nominal lattice parameter was set to a = 426 nm for both simulations and experiments  throughout the paper. 

\begin{figure}[H]
\includegraphics[height=18 cm]{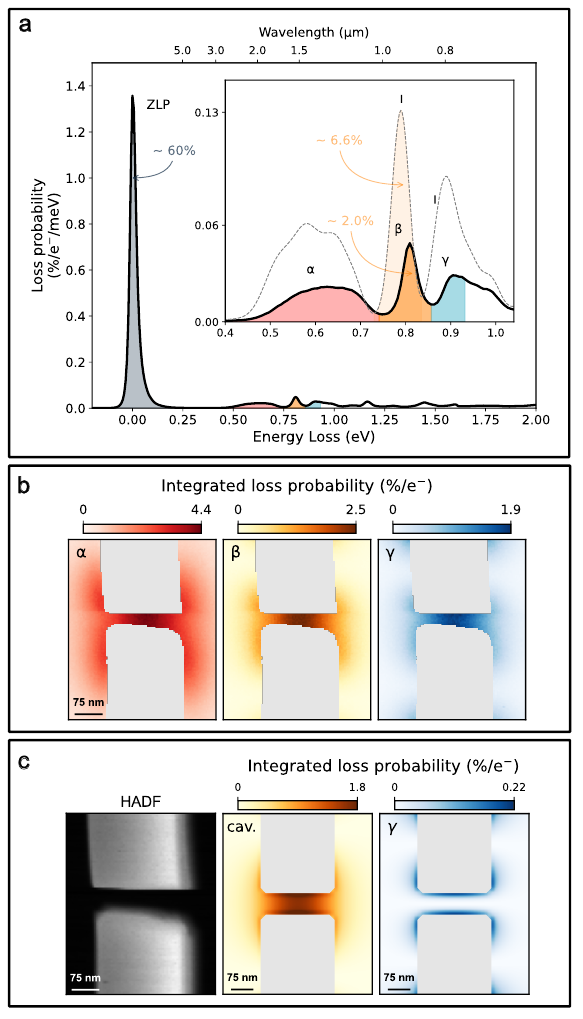}
\caption{\label{fig:par} \textbf{EELS spectral-imaging of a parallel configuration sample}  a. Spectrum taken with the beam passing through the center of the slot. The inset shows a magnified view of the spectrum around the main resonances as well as the corresponding FDTD simulated spectrum (dotted curve). The position of the cavity mode, as well as the mode in the $\gamma$ peak used to generate the maps in c. are indicated with a vertical mark. b. Maps of the $\alpha$, $\beta$ and $\gamma$ peaks, filtered on the energy windows indicated in the inset of a. c. Experimental HAADF and FDTD simulations. FDTD simulations have been performed for individual modes (cav.: cavity mode, and $\gamma$) indicated in the inset of the a., and therefore are less intense than the corresponding spectra that include several convoluted modes.}
\end{figure}

We show in Figure \ref{fig:par} experimental and simulation results for a parallel geometry  sample. The nominal $\rho$  is equal to 0.255. All designs with different $\rho$ values exhibit essentially the same sequence of modes but having different energies (see later in the manuscript). A spectrum measured with the beam centered in the middle of the slot is presented in Figure \ref{fig:par}a. A series of peaks can be directly observed, ranging from $\approx~50$ meV to a few eVs. We are interested in the energy range close to the band gap energy of the photonic-crystal, corresponding for the sample in Figure \ref{fig:par} to the energy-resolution-limited sharp peak $\beta$ (0.81 eV) and the broader peaks $\gamma$ (0.95 eV) and $\alpha$ (around 0.7 eV).  As shown in Figure \ref{fig:par}b, the intensity of the three  peaks are mostly localized in the very center of the structure. The measured probability of photon creation per incident electron is more than $2\%$ for the $\beta$ peak, i.e. around 1 creation of excitation every 50 incoming electron. As will be shown later, it can reach up to $3.2\%$ for other $\rho$ parameters.  This relatively strong coupling decreases rapidly as the beam is moved away from the slot.

The comparison with simulations is excellent in energy position of the different peaks and global spectral shape as shown in Figure \ref{fig:par}a. We also note that the absolute values given by the simulations, that do not rely on any fitting parameters, and experimental probabilities are sizable. Differences between experiments and simulations might rely in slight geometrical parameters differences, imperfect roundness of the holes, imperfect  knowledge of the silicon index and dispersion, and imperfect energy calibration of the EELS spectrometer.  Simulated spectra before convolution (Supplementary Figure S3) show that all experimental peaks are a sum of several ones that cannot be disentangled experimentally in energy position. The simulated $\beta$ peak is made up of only a restricted number of peaks. Amongst these peaks, a dominant mode - the cavity mode - can be distinguished because it is  several order of magnitudes more intense and with a quality factor typically two orders of magnitude larger (see supplementary Figure S3). 

Altogether, intensities and quality factors balance in such a way that the cavity mode represents typically 30 $\%$ of the total spectral weight of the $\beta$ peak.


The comparison with simulations is also remarkable for the $\gamma$ peak, both in shape and intensity in the spectrum. However, the absolute value in the simulated map, is  different from the experimental one. Also, the $y$ symmetry is different between experiments and simulations. In both cases, this is  again because only one specific mode has been simulated out of all those that are constituting the experimental peak.

At energies higher than around $1~eV$, there exists a plethora of modes that we will not discuss here as they fall above the band gap of Si (not to be confused with the band gap of the photonic-crystal itself). Typical spectra and filtered images are shown in the supplementary materials Figure S4 and S5.

\begin{figure}[H]
\includegraphics[width=15cm]{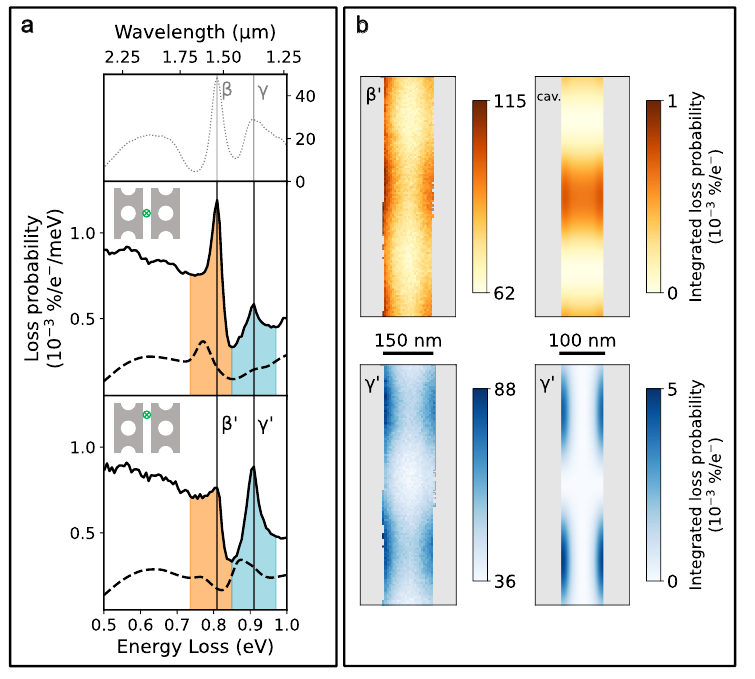}
\caption{\label{fig:perp} \textbf{EELS spectral-imaging of a perpendicular configuration sample} a. Spectra extracted from two representative positions indicated in the inset and highlighted in b, and compared to the spectrum of Figure \ref{fig:par}a at the top (nominal $a$ and $\rho$ are the same on both figures). The background due to the tail of the ZLP has been subtracted (see supplementary). The FDTD spectra are also shown as dashed lines. b.  Experimentally filtered maps around peak $\beta'$ and $\gamma'$ on the energy windows indicated in a. Single mode maps corresponding for the cavity mode and and one representative mode of the $\gamma$ bands. Intensities are substantially lower than the filtered maps due to a relatively even breakdown of spectral weight of the modes in the perpendicular direction. }
\end{figure}

The perpendicular geometry is exemplified in Figure \ref{fig:perp} for the same parameters as the parallel geometry of Figure \ref{fig:par}. Here again, several modes are observed in the same energy ranges (see supplementary Figure S6 for a full spectrum), but we focus on the energy region between 0.6 and 1 eV. Note that the tail of the ZLP has been removed by fitting a power-law, see supplementary Figure S7. In Figure \ref{fig:perp}a (middle, bottom), two main peaks are seen, around 0.8 eV (peak $\beta'$) and 0.9 eV (peak $\gamma'$). The energy position of these peaks match that of the $\beta$ and $\gamma$ peaks (see top of Figure \ref{fig:perp}a). The position of the beam is emphasized in the respective insets. From the energy filtered map in Figure \ref{fig:perp}b, we can see that their spatial distribution are aligned  (peak $\beta'$) and shifted by half a period (peak $\gamma'$) with the hole position. The intensity of the peaks is two orders of magnitude smaller than in the other geometry. The simulated spectra and maps in Figure \ref{fig:perp} fit very well in energy position and intensity. The absolute interaction probabilities are also sizable between experiment and theory in the spectra. However, the experimental and simulation maps show one to two orders of magnitude difference in intensity. This is because, contrary to the case of the parallel geometry, where only one mode contributes to $\sim$ 30 $\%$ of the intensity, here all modes contribute almost equally. Experimental maps are summed up over all modes while simulations retain only one. This emphasizes a key feature of our optimization scheme that has been developed for optimized coupling to the cavity mode in the parallel direction and not the perpendicular one.


The coupling between the electron and the modes of the structures can be tuned by phase matching. Usually, one tune the acceleration voltage for a given sample geometry. Unfortunately, the use of a monochromator does not allow for sufficient tuning range. An alternative method consists in changing systematically $\rho$, as it is well-known that the whole band diagram scales with the it \cite{Joannopoulos2008}.

\begin{figure}[H]
\hspace{-0.6cm}
\centering
\includegraphics[width=17cm]{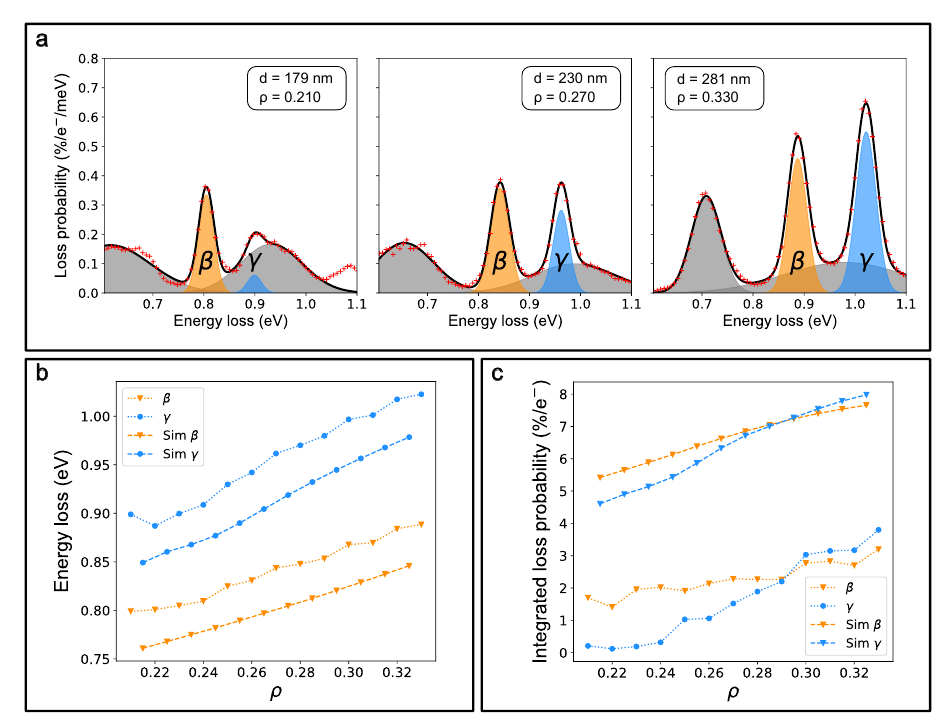}
\caption{\label{fig:hole} \textbf{Influence of the ratio $\rho$ on modes properties.} a. Spectra for three different holes diameters and same hole separation in the parallel configuration. Gaussian fitting is also shown, and parameters are indicated in the legend. b. $\rho$ dependence of the energy of the peak $\beta$ and $\gamma$ and comparison with FDTD simulations. c. $\rho$ dependence of the intensity for peak $\beta$ and $\gamma$ and comparison with simulations.}
\end{figure}

Therefore, as exemplified on Figure \ref{fig:hole} we have performed systematic measurements on parallel configuration samples as a function of $\rho$. 

$\rho$ values were changed from 0.210  to 0.330. Three representative spectra for increasing hole diameters are displayed in Figure \ref{fig:hole}a, showing an energy and intensity increase. These tendencies are synthesized on Figure \ref{fig:hole}b and c.  In Figure \ref{fig:hole}b, one can see the energy dependence of the $\beta$ and $\gamma$ peaks as a function of $\rho$ both for experimental and theoretical values. The slopes are in excellent agreement. There is only a slight and constant shift in the energy value. Figure \ref{fig:hole}c presents the intensity dependence of the modes $\beta$ and $\gamma$. Experiments and simulations results are in good agreement although simulations are constantly larger by  typical factor of 2 to 3. We want here again to emphasize that the simulations have  been done without any fitting parameters. We note that the maximum experimental coupling for the peak $\beta$ is shown to be 3.2$\%$ per incoming electron.

Before trying to identify the different modes, we need to comment on the effect of monochromation on the experiment. High monochromation leads to a decrease of the effect of the ZLP tail, making low energy resonances (down to the far-IR \cite{Li2021}) much easier to measure, but we must recognize the finite resolution of the STEM-EELS system used here, which was from $\approx 20$ to $\approx 40$ meV at 100 keV. This has to be compared to few physical quantities at hand here. First, there are few physical quantities that should be resolved: the bang gap of the photonic-crystal, of the order of 150 meV, the energy difference between dielectric and air slot modes, of the order of 70 meV, and the bandwidth of the bottom (dieletric) and top (air) photonic-crystal bands as seen in EELS.
Second, there are physical quantities that cannot be resolved, especially the linewidth of the cavity mode (of the order of a few $\mu$eV, the energy difference between the air slot modes and the top band of the photonic-crystal (typically less than 10 meV), and the energy difference between sub-modes of the slot (see Figure \ref{fig:design}).

We are now in the position to detail the link between  peaks measured in EELS and the main modes of the structures.

From their energy, the $\alpha$ feature corresponds to the bottom (dielectric) band of the photonic-crystal slab. The $\gamma$ mode is made up of a  band encompassing air modes related to the top  of the band structure of the PBG and the air slot modes, which are partly degenerated as seen on Figure \ref{fig:design}c. The $\beta$ peak can be associated to the even dielectric slot modes $D_e$. Beyond the correspondence between experimental and simulated EELS spatial distribution, they both more generally agree with the  $|E_x|^2$ distribution of $D_e$ and confirms our interpretation (see Supplementary Figure S2) - indeed, EELS is closely related to the electromagnetic density of state along the electron trajectory \cite{Abajo2008a}. Together with the already discussed argument that  the spectral weight of the cavity mode in the simulation which account for  30\% of the total intensity of the $\beta$ peak, this definitely proves that the $\beta$ peak is dominated by the coupling of the electron beam to the cavity mode.

If we now switch to the perpendicular case, we note that the $\beta'$ energy matches also that of the dielectric slot modes. Both experimental and simulated maps show maxima aligned to the holes centers, as expected for dielectric modes. Indeed, EELS roughly resembles  the corresponding $|E_z|^2$ field distribution (see supplementary Figure S2).  $|E_z|^2$ is mapped in the perpendicular geometry, and $|E_x|^2$ in the parallel one. Now, $|E_z|^2$ and $|E_x|^2$ are shifted by half a period in the slab plane, ie, while $|E_z|^2$ is maximum between two holes as confirmed by the EELS measurement (see Figure \ref{fig:perp}), the $|E_x|^2$ should peak at the hole positions (see Supplementary Figure S2). Same reasoning applies to $\gamma'$ which corresponds to the air band, possibly mixed with the air slot modes. This is also confirmed by EELS simulations. Finally, there is  a difference between experimental and simulated maps for $\beta'$ and the cavity mode. The simulation clearly shows an even mode behaviour (as selected), while the experimental has a mixed character. This is because, contrary to the parallel case, no special phase matching is expected in that geometry, therefore all the modes of the  $D_e$ and $D_o$ branches can contribute to the EELS signal. Beyond  the fast periodic modulation of period $a$, the lack of spectral resolution in EELS does not allow to image the envelopes of the different modes.

To summarize, the $\alpha$ peak  is related to the dielectric band of the photonic-crystal slab, while $\gamma$ and $\gamma'$ correspond to its air band and/or the air quasi-TE guided mode of the slot. The $\beta$ and $\beta'$ are both made up of dielectric slots modes. However, while $\beta'$ is made up of a mixture of even and odd modes, $\beta$ is constituted by a finite number of even modes, and dominated by the contribution of the cavity mode.

Finally,  the achieved  coupling strength, leading to the creation of few percents of photons  per electrons in a single mode, narrow band cavity, is relatively high. Indeed, reported experiments in EELS, cathodoluminescence or PINEM have largely focused the study of plasmons in nanoparticles, for which a typical value for coupling is an order of magnitude lower. This, despite the fact that the plasmons have extremely high density of states. In our case, both the length and access to the phase-matching explains the coupling values. Much larger coupling values have been reported \cite{Adiv2023} for the EELS of surface-plasmon polaritons of surfaces probed in grazing incidence, but again in this case the density of states is enormous compared to the present case. Our situation is closer to that of Feist et al. \cite{Feist2022}, who reported a slitghly larger  coupling efficiency (around 2.5 $\%$) to a single mode of an optical micro-resonator. The main difference is the modal volume, which can be very roughly estimated to be $\sim$ 260 $\lambda^3$. This means that our design provides similar coupling but with a modal volume which is typically 3 orders of magnitude lower, therefore an interesting candidate for quantum optics applications.


In conclusion, we have developed photonic band gap based cavities compatible with EELS experiments in a TEM. Spatially resolved EELS experiments combined to FDTD simulations made it possible to prove that, despite not having enough spectral resolution to resolve cavity modes, those could be isolated within the band gap and mapped. We have shown that coupling of slots modes can be as high as 3.2 $\%$ of photons created per incoming electrons, out of which a third are directed to the cavity mode with a quality factor Q of the order of $2.5.10^5$.
Resolving spectrally the cavity modes could be done in the future using PINEM or electron energy gain spectroscopy \cite{Henke2021,Auad2023}.
The coupling between the electron and the cavity is already sufficiently high that we can hope for use in experiments requiring high-Q cavities \cite{Kfir2019,kfir2020,Wang2020,DiGiulio2019}. It could be further improved by a better phase-matching - which would require a more systematic design and production of samples - and workarounds for charging issues.  
\section*{Acknowledgments}
This project has been funded in part by the European Union through the Horizon 2020 Research and Innovation Program (grant agreement No. 101017720 (EBEAM)), the French National Agency for Research under the program of future investment TEMPOS-CHROMATEM (reference No. ANR-10-EQPX-50), QUENOT (ANR-20-CE30-0033) and ANR OFELIA (ANR-21-CE24-007). This work was supported by the French RENATECH network. MK thanks Javier Garcia de Abajo for fruitful discussions on the topic.

\section*{Methods}

\subsubsection*{Simulations}
Band diagrams and the corresponding mode patterns of the waveguide were calculated by  FDTD with MEEP software \cite{Oskooi2010}. A single period of the photonic crystal is simulated using Bloch boundary condition in the direction of the waveguide and perfectly matched layers (PML) in the two other directions. 

   The EELS spectrum $\Gamma(\hbar\omega,y,z)$ for a given position y, z, or equivalently the probability to create a photon in the cavity per electron and unit of energy can  be calculated by the time-domain Fourier transform of the field with proper prefactors: 
\begin{equation}
\Gamma(\omega,y,z)=\frac{ev}{\pi\hbar\omega}\textrm{Re}\int_{-\infty}^{+\infty}\hat{E}_{x}^{\textrm{ind}}(vt,y,z,\omega)e^{i\omega t}dt
\label{eq:EELS}
\end{equation}
with $\hat{E}_{x}^{\textrm{ind}}(x,y,z,\omega)=\int_{-\infty}^{+\infty}E_{x}^{\textrm{ind}}(x,y,z,t)e^{-i\omega t}dt$, the Fourier transform of the induced electric field and $v$ the speed of the electron.

For the simulation of EELS spectra, a digital analog of the continuous integral in eq. \ref{eq:EELS}  is calculated using an home-made finite-difference in time-domain (FDTD) software \cite{Oskooi2010,Song2021}. The sample was modelled using the same parameters as the experimental ones, with the silicon refractive index set to 3.45 and a FDTD resolution of 18 nm. An electron travelling along the slot is modelled by a single FDTD pixel current source travelling with a velocity corresponding to an energy of 100 keV as in the EELS experiment. The field is recorded at every FDTD time step and at every spatial pixel along the whole path taken by the electron (a straight line). From the resulting field,  the field generated by the crossing of a single electron in an empty simulation domain is subtracted to remove the transients generated by the entrance and exit of the charged particle in the simulation domain. Then, at each spatial pixel, a temporal Fourier transform is calculated using a Padé approximant technique to increase the accuracy of the digital Fourier transform. Finally a digital analog of the continuous integral in eq. (1) is calculated.

This method requires one simulation per pixel in the (y,z) plane perpendicular to the electron trajectory and thus is not well adapted to simulate the spatial field mapping achieved in EELS experiment. A faster calculation has been done for a particular mode since its electric field pattern, $\mathbf{u}_{k}(\mathbf{r},\omega_{k})$, with $\omega_{k}$ the complex resonant pulsation of the cavity, can be calculated by FDTD quickly. Indeed, the induced field can be calculated with the Green tensor, $\overleftrightarrow{\mathbf{G}}(\mathbf{r},\mathbf{r}',\omega)$ of the cavity,  
\begin{equation}
 \mathbf{E}^{\textrm{ind}}(\mathbf{r}(t),\omega)=-i\omega\int_{\mathbb{R}^{3}}\overleftrightarrow{\mathbf{G}}(\mathbf{r},\mathbf{r}',\omega)\mu_{0}\mathbf{j}(\mathbf{r}'(t),\omega)d^{3}\mathbf{r}'
 \end{equation}

with $\mathbf{j}(\mathbf{r}'(t),\omega)=e\frac{\overrightarrow{v}}{|v|}e^{-i\omega x'/v}\delta(y_{0},z_{0})$, the current density generated by the crossing of a single electron travelling along $x$. Decomposing the Green tensor on the basis of the cavity modes, we get for an electron travelling in the slot along the x direction 
\begin{equation}
\Gamma(\omega,y,z)=\sum_{k}\frac{e^{2}}{2\pi\hbar\omega}\frac{1}{\mathcal{E}_{0}}\textrm{Re}\left(\frac{i\omega}{\omega_{k}^{2}-\omega^{2}}\right)\left| FT[u_{xk}(x',y_{0},z_{0},\omega_{k})]\right|_{-\frac{\omega}{v}}^{2}
\end{equation}

 where the $\mathbf{u}_{k}(\mathbf{r},\omega_{k})$ have been normalized so that $\frac{1}{2}\int_{\mathbb{R}^{3}}\varepsilon_{0}\varepsilon_{r}(\mathbf{r}')\boldsymbol{u}_{k'}^{*T}(\mathbf{r}',\omega_{k})\mathbf{u}_{k}(\mathbf{r},\omega_{k})d^{3}\mathbf{r}=\mathcal{E}_{0}\delta_{k'k}$ with $\mathcal{E}_{0}=1$ Joule and $ FT[u_{xk}(x',y_{0},z_{0},\omega_{k})]_{\frac{\omega}{v}}$ is the spatial Fourier transform of the mode along the x direction evaluated at a wave vector $\frac{\omega}{v}$.
 Similar simulations have been performed for the perpendicular case, but considering $E_{z}(x,y,vt,\omega$) instead of $E_{x}(vt,y,z,\omega)$.

Spectra were convoluted by a 40 meV gaussian to account for experimental spectral broadening.

\subsubsection*{Sample fabrication}

To produce the samples, a silicon on insulator wafer with a 220-nm top layer on 2-\textmu m burried oxide was spin-coated with ZEP electronic resist. Photonic structures were lithographied with a Raith ebpg 5200 system at 100 keV. Patterns were transfered into the silicon layer by inductively coupled reactive ion etching (ICP-RIE) using a SF$_{6}$-C$_{4}$F$_{8}$ gas mixture. The resist was removed with a solvent. The samples are then saw diced to a size of 250-\textmu m $\times$ 2.8 mm compatible with TEM holders and wet etched by hydrofluoric acid to remove the buried oxide.

For the perpendicular geometry experiments, a backside lithography aligned with the photonic crystal and a subsequent etching has been performed before dicing the sample in order to remove the substrate locally.

\subsubsection*{Experiments and data analysis}
EELS experiments have been performed on a NION Hermes microscope (CHROMATEM) at 100 keV, fitted with a Merlin Quantum Detector camera.  The spectral resolution was  set to 30-40 meV (as measured on the ZLP) with a dispersion of 6.7 meV/channel. Incident semi-angle was 5 mrd and acceptance semi-angle 30 mrd. Spectral-images with typically $10^4$ pixels were recorded with a typical 30 ms dwell time for the parallel direction and 100 ms for the perpendicular one. The spectra were normalized by their total integrated number of counts. For an electron in vacuum, this ensures that the sum of all events probability is one, as expected as all the electrons are detected on the spectrometer camera. As this assumption is wrong for electrons travelling in the bulk, corresponding pixels have been masked in the data presentation. Then, all spectra were normalized by the dispersion and represented in percentage. At the end, the spectra amplitude are represented  as the ($\%$) probability to create an excitation in the material per unit of energy. Fitted or integrated maps are then directly showing this probability at each pixel. All data were analyzed and processed using Hyperspy \cite{francisco_de_la_pena_2022_7263263}.

\bibliography{HighQCavityEELS}
\end{document}